**Do readers use character information when programming return-sweep saccades?**

Martin R. Vasilev[1] *

Victoria I. Adedeji[1]

Calvin Laursen[1]

Marcin Budka[2]

Timothy J. Slattery[1]

[1]Bournemouth University, Department of Psychology, United Kingdom

[2]Bournemouth University, Department of Computing and Informatics, United Kingdom

Corresponding author* at:

Department of Psychology, Bournemouth University

Poole House, Talbot Campus, Fern Barrow

Poole, Dorset, BH12 5BB, United Kingdom

Phone: +44 7835202606

Email: mvasilev@bournemouth.ac.uk

**Word count:** 7800 (excluding Tables, Figures, and References)

*Note*: color should not be used for any figures in print.

Abstract

Reading saccades that occur within a single line of text are guided by the size of letters. However, readers occasionally need to make longer saccades (known as return-sweeps) that take their eyes from the end of one line of text to the beginning of the next. In this study, we tested whether return-sweep saccades are also guided by font size information and whether this guidance depends on visual acuity of the return-sweep target area. To do this, we manipulated the font size of letters (0.29 vs 0.39º per character) and the length of the first line of text (16 vs 26º). The larger font resulted in return-sweeps that landed further to the right of the line start and in a reduction of under-sweeps compared to the smaller font. This suggests that font size information is used when programming return-sweeps. Return-sweeps in the longer line condition landed further to the right of the line start and the proportion of under-sweeps increased compared to the short line condition. This likely reflects an increase in saccadic undershoot error with the increase in intended saccade size. Critically, there was no interaction between font size and line length. This suggests that when programming return-sweeps, the use of font size information does not depend on visual acuity at the saccade target. Instead, it appears that readers rely on global typographic properties of the text in order to maintain an optimal number of characters to the left of their first fixation on a new line.

Word count: 248 words

# 1. Introduction

Return-sweeps are the largest saccades during reading. Their function is to move gaze from the end of one line of text to the beginning of the next (Rayner, 2009). While return-sweeps are common in everyday reading, their planning is not well understood as most eye-movement studies of reading have used single-line sentences where return-sweeps are absent. Consequently, it is unclear whether return-sweeps and saccades that occur within a single line (i.e., intra-line saccades) are guided by the same oculomotor principles. For example, intra-line saccades are guided in part by the number of characters they travel rather than a specific distance in degrees of visual angle (Morrison & Rayner, 1981; O'Regan, 1983). However, as return-sweeps traverse a much larger distance, it is not clear if return-sweeps are guided by character information like intra-line saccades, and whether the use of such information is modulated by visual acuity constraints. We suspect that return-sweep guidance may be distinct from the guidance of intra-line saccades.

## 1.1. Intra-line Saccades

Most reading saccades are intra-line, as they begin and end within the same line of text. Intra-line saccades are usually about 7-8 characters long (Rayner, 1978; Yang & Vitu, 2007). In intra-line reading, the initial fixation on words usually falls near the word's centre and shifts slightly further to the left of the centre with increasing word length (McConkie et al., 1988; Rayner, 1979; Vitu et al., 1990). This is known of as the *preferred viewing location* (PVL) effect (Rayner, 1979). Additionally, initial landing positions are influenced by *launch site*— that is, the distance between the position of the previous fixation and the start of the word on which the saccade landed. More specifically, saccades launched from a short distance tend to overshoot the word's centre, whereas saccades launched from a long distance tend to undershoot it (e.g., McConkie et al., 1988; Radach & McConkie, 1998).

McConkie et al. (1988) argued that the PVL and launch site effects can be explained by a few principles: 1) Each word object has a "functional target" that readers aim for— this target is assumed to be the centre of the word (also known as the *optimal viewing position* [OVP]; O'Regan, 1992; O'Regan & Levy-Schoen, 1987; Rayner, 2009); 2) There is *saccadic range error* (SRE; Kapoula, 1985; Kapoula & Robinson, 1986)[1], or a central-tendency bias towards making saccades that correspond to the centre of all possible saccade lengths within a given task/ set, such that the eyes tend to overshoot the target when it is close by and undershoot it when it is far away (with the centre corresponding to accurate saccades where undershoots and overshoots are equalized); 3) The SRE is modulated by the duration of the previous fixation, with longer fixation durations attenuating its effect; and 4) Saccades are also influenced by random error (i.e., variation of landing sites around the mean), which increases with greater saccade length. Thus, the combination of a functional target and the presence of the SRE allows McConkie et al.'s (1988) model to explain the modulation of landing sites by launch site.

McConkie et al.'s (1988) key principles have since been implemented in many recent models of eye-movement control during reading (e.g., Engbert, Nuthmann, Richter, & Kliegl, 2005; Reichle, Warren, & McConnell, 2009; Snell, van Leipsig, Grainger, & Meeter, 2018), which assume that intra-line saccade targeting is word-based. However, a non-word-based "Center-of-Gravity" theory has been proposed (Vitu, 2008, 2011), which has also received empirical support (Albrengues et al., 2019; Vitu, 1991; Yao-N'Dré et al., 2014). In this theory, eye guidance mainly reflects the integration of bottom-up, luminance-contrast signals that take the eyes towards a centre-of-gravity configuration of letters, regardless of which word these letters belong to (Albrengues et al., 2019).

---

[1] It is should be noted that more recent evidence that has failed to support the existence of the SRE effect (Gillen et al., 2013; Nuthmann et al., 2016), although its presence may depend on the availability of learned prior knowledge in the task (Krügel et al., 2020).

Research has shown that saccade length in alphabetical languages is guided more by the number of letters that the eyes travel than by simple distance in visual angle. For example, Morrison and Rayner (1981) changed the viewing distance of the text so that the width of each letter was 0.35º, 0.47º or 0.69º. They found that the amplitude of intra-line saccades remained the same in terms of the number of characters traversed. This finding was later replicated by O'Regan (1983) using a similar method. These results suggest that readers adjust the absolute size of their saccades in visual angle to match the size of letters in the text.

Studies that directly manipulated the font size of text have largely confirmed these results (but see Shu, Zhou, Yan, & Kliegl, 2011; Yan, Zhou, Shu, & Kliegl, 2015 for potential differences in non-alphabetic scripts). For instance, Bullimore and Bailey (1995) found that the forward saccade length in letters remained relatively constant with increasing font size when subjects with normal vision read a text chart. Similarly, Beymer et al. (2008) used a Verdana font that ranged from 10 to 14 pt. in size. They also found that saccade length in letters was not influenced by font size. Additionally, Miellet et al. (2009) reported that parafoveal magnification of the upcoming text did not influence saccade length in letters compared to reading without magnification, which also suggests that intra-line saccades are guided by character information. Saccade length has also been shown to scale with the amount of spacing used between letters and words, even when the letters themselves remain the same size (Slattery et al., 2016; Slattery & Rayner, 2013).

However, there is some evidence suggesting that saccade length in letters could be mildly influenced by font size. Franken, Podlesek, and Možina (2015) found that saccade length in letters decreased with increasing font size. In their study, letter sizes ranged from 0.21º to 0.46º and the biggest change in saccade length was observed for the smallest font sizes. Additionally, Yan et al. (2015, Experiment 2) manipulated the font size of German sentences where the letter width was 0.30°, 0.45°, or 0.60°. They also found that forward saccade length in letters decreased with increasing font size (the largest difference was about a character). Furthermore, Yao-N'Dré, et al. (2014) found that character size in French can impact both the preferred viewing location and launch site effects in a task where participants had to read a list of words and search for an animal name. Thus, Yao-N'Dré et al.'s (2014) findings may pose a challenge to reading models that assume word-based saccade targeting. In summary, while recent studies (Franken et al., 2015; Yan et al., 2015; Yao-N'Dré et al., 2014) suggest that it may be an oversimplification to conclude that saccade length in letters is completely independent from font size, the bulk of the evidence suggests that intra-line saccades depend, at least partially, on font size information.

## 1.2. Return-sweep Saccades

Return-sweeps move gaze from the end of one line of text to the beginning of the next. They travel much farther than intra-line saccades, typically some 30-70 characters. Return-sweeps are usually launched 4-6 characters from the end of the previous line (Hofmeister et al., 1999; Parker, Slattery, et al., 2019) and typically land 5-8 characters from the beginning of the new line (Hofmeister, 1998; Parker, Slattery, et al., 2019; Slattery & Vasilev, 2019). One important factor that influences return-sweeps is line length. With longer lines, readers tend to land further to the right from the line start (Hofmeister et al., 1999; see also Parker, Nikolova, Slattery, Liversedge, & Kirkby, 2019). This rightward shift in landing positions likely results from increased saccadic undershoot error, as longer saccades are more likely to undershoot their target near the left margin (Bartz, 1967; Henson, 1979). Of course, this undershoot explanation works only if one assumes that there is an explicit target at the start of the new line.

There are at least two possible explanations for return-sweep undershoot errors. Saccadic undershoot may be a basic aspect of the oculomotor system as long saccades are often followed by a shorter corrective regression (Becker, 1972). For example, Becker and Fuchs (1969) noted that long saccades typically travel about 90% of the distance to the target and that a second, corrective saccade usually covers the remaining 10%. Most return-sweeps are long saccades and often fail to reach the beginning of the new line at once (Andriessen & de Voogd, 1973). Rather, readers generate a secondary saccade after approximately 40-60% of all return-sweeps (Hofmeister, 1998; Parker, Slattery, et al., 2019; Slattery & Vasilev, 2019). In research on return-sweeps, these secondary saccades have traditionally been called corrective saccades (Abrams & Zuber, 1972; Heller, 1982; Hofmeister, 1998; Hofmeister et al., 1999; Radach et al., 1999). This terminology is supported by the finding of Hofmeister's (1998) Experiment 1 that the amplitude of the "corrective saccade" was negatively correlated with the duration of the preceding fixation[2]. We will refer to them here as corrective regressions to denote the fact that they move gaze to an earlier point in the text.

However, saccadic undershoot errors that occur in tasks where the stimuli consist of complex visual arrays may also be explained by the centre-of-gravity effect (Findlay, 1982; Findlay & Gilchrist, 1997; for reviews, see Findlay & Walker, 1999; Vitu, 2008). The centre-of-gravity effect would suggest that readers, targeting a location near the left margin, will land short of the start of the line because the text stimuli lie to the right of the margin and there are no stimuli left of the margin to help balance the center-of-gravity closer to the margin itself. However, these are not mutually exclusive explanations. The fact that corrective regressions are more likely to occur with long compared to short previous lines (Hofmeister et al., 1999) seems more parsimoniously explained by saccadic undershoot bias as the probability of making a corrective regression increases with greater target distance (Henson, 1979; Hyde, 1959; Weber & Daroff, 1971). However, the finding that return-sweeps land closer to the margin when the first word on the line is presented in bold (Slattery & Vasilev, 2019) is more easily explained by centre-of-gravity effects which can be influenced by the physical properties of stimuli such as luminance (Beauvillain et al., 1996). So, both undershoot bias and center-of-gravity effects may be required to explain the pattern of return-sweep undershoot errors.

---

[2] In a post-hoc analysis of the current experimental data, we also found a significant negative correlation between the corrective saccade amplitude and the prior fixation duration, $r(3721)= -0.249$, p<0.001, 95% CI [-0.279, -0.219].

The fixation between a return-sweep and a corrective regression has been called an *under-sweep fixation* by Parker, Kirkby, and Slattery (2017). These fixations are usually much shorter than the average reading fixation and last for approximately 120-160 ms (Hofmeister, 1998; Parker, Slattery, et al., 2019; Slattery & Parker, 2019; Slattery & Vasilev, 2019). Under-sweep fixations were originally thought to be unrelated to text processing and reflect the time needed to program a corrective regression (e.g., Abrams & Zuber, 1972). However, recent evidence has suggested that readers may acquire useful information during such fixations that can later aid reading (Slattery & Parker, 2019). Additionally, reducing the oculomotor error associated with return-sweeps does not improve reading speed (Slattery & Vasilev, 2019), further suggesting that readers acquire useful information during under-sweep fixations.

While the basic characteristics of return-sweeps are well documented, less is known about what information is used to program such saccades. Intra-line saccades are typically thought to target words' OVP (McConkie et al., 1988), which is facilitated by the fact that the next word usually falls in parafoveal vision. However, return-sweeps are much longer, and their target usually falls well into peripheral vision where acuity is limited. To test whether return-sweeps target the OVP of words similar to intra-line saccades, Slattery and Vasilev (2019) examined how landing positions are influenced by the length of the first word on the line.

They found that return-sweep landing positions were not influenced by line-initial word length, even when this word was formatted in bold to make it more prominent. Rather, readers appeared to target the left margin of the new line as landing positions shifted closer to the line start due to the bolding. This suggests that readers rely on different information for targeting return-sweeps than for targeting intra-line saccades.

As noted previously, intra-line saccades depend at least partially on font size information. However, there is surprisingly little evidence on how font size influences return-sweeps. To our knowledge, Hofmeister's (1998) Experiment 2 is the only study to address this question. In this experiment, eight participants read a text in four font size conditions that corresponded to letter widths of 0.27, 0.33, 0.39, and 0.44º. Hofmeister found that return-sweep landing positions in visual angle shifted further to the right of the line start with increasing font size. Because there are fewer characters to occupy the same physical area with larger fonts, this rightward shift suggests that return-sweeps may also be influenced by letter information. Additionally, corrective regression probability decreased with greater font size. However, the words at the beginnings and endings of lines were not controlled in Hofmeister (1998). This leaves open the possibility that lexical information around the launch and landing sites of return-sweeps was at least partially responsible for the reported effects.

### 1.3. Present Study

The present study manipulated character size to test whether return-sweep saccades are guided by character information in a similar way to intra-line saccades, and whether the use of character information depends on visual acuity constraints. This is important as return-sweeps differ from intra-line saccades in ways that could make character information a less reliable targeting cue. First, return-sweep targets usually fall well into peripheral vision and readers do not have the benefit of parafoveally previewing them (Parker, Nikolova, et al., 2019; Parker, Slattery, et al., 2019). Second, readers appear to target an area relative to the left margin rather than the centre of the first word on the next line (Slattery & Vasilev, 2019). Therefore, one plausible saccade targeting strategy would be to ignore global text characteristics and instead focus only on locating the left text margin. This strategy predicts that font size would not influence return-sweep landing positions in visual angle as the left margin remains constant across different font sizes.

A second possibility is that return-sweeps are targeted to place gaze a number of characters to the right of the margin for optimal visual encoding. In this scenario, readers must use global text characteristics such as font size when programming return-sweeps. Therefore, consistent with Hofmeister's (1998) results, their landing positions in visual angle should shift to the right with larger fonts so that gaze would start at a similar character across different font sizes.

This explanation bears some resemblance to O'Regan's (1990) "strategy-tactics" theory in which the eyes are guided by a general strategy of scanning the text based on its gross visual characteristics. Font size is one such characteristic that can influence saccadic behaviour and this could explain why return-sweep saccades in visual angle (but not necessarily in number of characters) may change with larger letters, as this would allow readers to scan the text more efficiently by landing in a more optimal position on the new line. Of course, both O'Regan's (1990) theory and more recent reading models (e.g., Engbert et al., 2005; Reichle, Pollatsek, Fisher, & Rayner, 1998) assume that saccades are targetted towards the centre of word-based objects, which appears to be inconsistent with the available evidence on return-sweep targetting (Slattery & Vasilev, 2019). Nevertheless, in light of Hofmeister's (1998) results, this still remains a viable hypothesis. Note that we test these predictions using landing sites measured in visual angle rather than number of characters as the number of character metric may result in a null prediction. That is, if readers adjust their oculomotor behaviour to attempt to maintain an optimal number of characters to the left of their landing site then there should be sizeable differences in these landing sites when measured in visual angle but not necessarily when measured in number of letters.

To examine the role of visual acuity, we also manipulated the length of the line from which readers launch their return-sweeps. With longer lines, visual acuity of the saccade target will be reduced as the target will move further into peripheral vision thus increasing visual crowding (Whitney & Levi, 2011). If return-sweeps are indeed guided by character information at the return-sweep target location, landing positions may be more strongly influenced by font size with shorter lines as the targets will be closer to the fovea than with long lines. Therefore, if the use of character information depends on visual acuity, the font size effect in landing positions should become smaller with longer lines. Alternatively, font size information obtained foveally, rather than at the peripheral target location, may be extrapolated to allow readers to adjust their return-sweep behaviour. If this were the case, we would expect a main effect of font size but no interaction with line-length[3].

Since corrective regressions following return-sweeps are more likely with long than short lines, this manipulation of line-length also affords us the opportunity to examine how font size information influences their probability (i.e. the likelihood that a return-sweep is followed by a corrective regression). Hofmeister (1998) suggested that such corrective regressions occurred after return-sweeps based on the number of letters to the left of the return-sweep landing position. As return-sweep landing position shifts rightward in terms of visual angle, the difference in number of characters to the left of this position between a small and a large font line will increase. For instance, given the size of the fonts used in the current study, a return-sweep that lands 1 degree of visual angle from the left margin would have 3.39 or 2.54 characters to the left of it in the small and large font conditions respectively. However, a return-sweep that lands 2 degrees from the left margin would have 6.78 or 5.08 characters to the left of it, respectively. Therefore, as landing positions in visual angle shift rightward with longer lines, the number of characters to the left of the landing site will increase more for the small compared to the large font size condition. Therefore, if Hofmeister's claim is correct, we may find an interaction between font size and line length with more corrective saccades in the small font condition which should be more pronounced with long lines. However, it is important to keep in mind that this prediction will depend on the distribution of landing sites across the conditions. As mentioned, it is possible that readers will adjust their return-sweep targeting so that they land on the same character in the different font size conditions which would invalidate this prediction.

---

[3] We thank an anonymous reviewer for pointing out this alternative.

To test these questions, a 2 (font size: small [0.29º] vs. large [0.39º]) x 2 (line length: short [16º] vs. long [26º]) within-subject design was used. This manipulation ensures that, for a given line length, the line will subtend the same visual angle for both font sizes, but it will have fewer characters with the large font than with the small font. The line length conditions were chosen as to create a sizeable difference in visual acuity and the distance to the saccade target. The full preregistered hypotheses can be read at https://osf.io/9sngw. In addition, we explored whether readers gradually learn to adapt their return-sweep targeting decisions based on exposure to a certain font size. This was done by examining how trial number within each font size block influenced landing positions. We expected that when readers switch from the smaller to the larger font, they will gradually learn to shift their landing position further to the right as the letters will be larger and occupy a greater area. Conversely, we expected the opposite trend to occur for participants starting with the large font block and moving to the small font block.

## 2. Method

## 2.1. Participants

Sixty-four Bournemouth University students (52 female) participated for course credit. Their average age was 20.05 years ($SD$= 1.42 years; range 18-26 years). All participants were naïve as to the purpose of the experiment and were fluent English readers who reported normal or corrected-to-normal vision and no history of reading disorders. Sixty-one participants were native English speakers while three participants were fluent readers who had used the language for at least five years. The study was approved by the Bournemouth University Research Ethics Committee (ID 25619). Each participant provided informed written consent. The study protocol was pre-registered before data collection (https://osf.io/9sngw).

The sample size was calculated *a priori* based on a power analysis using the PANGEA software (Westfall, 2015). The expected effect sizes were calculated based on Hofmeister's (1998) Experiments 1-2. The analysis indicated that at an alpha level of 0.05, 64 participants were needed to achieve 80% power (Cohen, 1988) of detecting the smallest effect size ($d$= 0.325).

## 2.2. Materials and Designs

The stimuli consisted of 100 declarative sentences (see Figure 1). Each sentence appeared on two lines. The experiment had a 2 x 2 within-subject design with *font size* (small vs large) and *line length* (short vs long) as the factors. In the small font condition, the width of all characters was 12 pixels (0.295º). In the large font condition, the width of all characters was 16 pixels (0.394º). The first line of text was 16º in the short line condition and 26º in the long line condition. When constructing each item, a maximum deviance of ±0.5º was allowed in both line length conditions; as such, the average line length across all items was 15.97º for the short line condition and 26.02º for the long line condition. While research shows that reading fonts of this size at these retinal eccentricities is not possible, reading rate was higher at 15º eccentricity than at 20º eccentricity across a range of large font sizes (Chung et al., 1998). Therefore, it is safe to assume that our line-length manipulation constituted a strong manipulation of visual acuity at the return-sweep target location.

The two independent variables were manipulated by changing the number of letters on the first line. Care was taken to ensure similar sentence meanings across the four conditions. The first and last four words on the first line were held constant across the four conditions. This ensured that readers processed the same words when they were about to make a return-sweep. The first line of text contained on average 54 characters in the small-font/ short-line condition (7 to 13 words; $M$= 10.12 words), 41 characters in the large-font/ short-line condition (6 to 10 words; $M$= 7.91 words), 88 characters in the small-font/ long-line condition (11 to 21 words; $M$= 15.71 words) and 66 characters in the large-font/ long-line condition (9 to 16 words; $M$= 12.26 words). For each item, the second line was identical in all experimental conditions and contained on average 50 characters (5 to 14 words; $M$= 8.67 words). The assignment of conditions to sentences was Latin-square counter-balanced across participants. This meant that each participant saw each item only once (in one of the conditions), but that each item appeared equally often in all conditions across all participants. Therefore, participants were exposed to any given item only once and could not learn to predict the text on the second line as this text was unique for every item. The two font size conditions were blocked, and block order was counter-balanced across participants. The items within each block appeared in a pseudo-random order. The line length manipulation was not blocked and thus the items appeared pseudo-randomly within each font size block.

**a) small-font/ short-line**

> The three musicians organised a tour last year to meet
>
> their fans and celebrate the release of their new studio album.

**b) large-font/ short-line**

> The two DJs did a tour last year to meet
>
> their fans and celebrate the release of their new studio album.

**c) small-font/ long-line**

> The three award-winning musicians decided to organise a national tour last year to meet
>
> their fans and celebrate the release of their new studio album.

**d) large-font/ long-line**

> The award-winning DJs organised a national tour last year to meet
>
> their fans and celebrate the release of their new studio album.

*Figure 1.* An example sentence used in the four experimental conditions. The line length manipulation occurred only on the first line.

## 2.3. Apparatus

Eye-movements were recorded with an SR Research EyeLink 1000 eye-tracker at 1000 Hz using a tower-mount set-up. The average horizontal and vertical accuracy of the system was 0.25 - 0.5°. The calibration offset error calculated at validation did not differ between the horizontal (*M*= 0.299º; *SD*= 0.104º) and vertical axes (*M*= 0.314º; *SD*= 0.116º), *t*(124.6)= -0.789, *p*= 0.431, thus suggesting that the error was comparable in both dimensions. Viewing was binocular, but only the right eye was recorded[4]. While a bite bar was not used, participants' head was stabilised with a chin-and-forehead rest. The text was presented on a Cambridge

---

[4] The left eye was recorded for two participants due to tracking problems caused by wearing glasses or contact lenses.

Research Systems LCD++ monitor (resolution: 1920 x 1080 pixels; refresh rate: 120 Hz). The text was formatted in a monospaced Consolas font and appeared as left-aligned black letters over white background. The stimuli were centred vertically and appeared with a 200-pixel offset horizontally with double-spaced lines. The distance between participants' eye and the monitor was 80 cm. At this distance, each letter subtended 0.295º horizontally in the small font condition and 0.394º in the large font condition. Because visual angle rather than characters was used as the depended measure, the accuracy of the eye-tracker did not preferentially affect one font size condition over the other. The distance between the two lines of text was 1.08º in the small font and 1.28º in the large font condition. The experiment was programmed in Matlab R2014a (MathWorks, 2014) using the Psychtoolbox v.3.0.11 (Brainard, 1997; Pelli, 1997) and Eyelink (Cornelissen et al., 2002) libraries. The experiment was run on a Windows 7 PC.

## 2.4. Procedure

A 9-point calibration was performed before the experiment. Calibration accuracy was monitored with a drift check before each trial. Participants were recalibrated whenever the error was > 0.4º. The experiment started with six practice trials (three in the small font and three in the large font condition). Each trial started with a black gaze-box centred at the first letter in the sentence. Once the gaze-box was fixated, it disappeared, and the sentence was presented on the screen.

Participants clicked the left button of the mouse to indicate they had finished reading the sentence. After 40% of trials, a True/ False comprehension question was presented, and participants used the mouse to select the correct answer. For example, in the sentence "The three musicians organised a tour last year to meet their fans and celebrate the release of their new studio album.", the question was "The artists celebrated the release of a new album. True/False?". The questions could be answered equally well in all conditions as they were based on information that was shared among them. The experiment lasted about 25-35 minutes and participants took a short break halfway through.

## 2.5. Data Analysis

Two main measures were analysed: 1) landing position of return-sweeps in visual angle relative to the start of the line (continuous measure); and 2) the probability of making an under-sweep fixation (binomial measure). An *under-sweep* was defined as a return-sweep saccade that undershoots the line start and is followed by a leftward saccade (Parker et al., 2017). Additionally, an analysis of intra-line and return-sweep saccade length (continuous measure) is presented in the Supplemental materials. The data were analysed with (Generalised) Linear Mixed Models ((G)LMMs) using the lme4 package v.1.1-21 (Bates et al., 2014) in the R software v.3.53 (R Core Team, 2019). Sum contrast coding was used for the font size (small font: -1; large font: 1) and line length factors (short line: -1; long line: 1). With sum contrast coding, the intercept in the model indicates the grand mean of all conditions and the slope estimate indicates the difference from the grand mean for the condition coded as "1". In the landing position and under-sweep probability models, return-sweep launch position was added as a covariate. Return-sweep launch position (continuous measure) was calculated as the distance in visual angle from the end of the first line. To improve the scaling of the models, return-sweep launch position was centred with a mean of 0. No other variables were transformed in the models.

Random intercepts were added for both participants and items (Baayen et al., 2008). As indicated in the pre-registration, we planned to add random slopes for font size and line length for both participants and items if the models converged (Barr et al., 2013). If they did not, we planned to remove slopes until convergence was achieved. The landing position and under-sweep probability models converged only with a random slope of line length for participants and items. The saccade length model converged only with a random slope of font size for participants. The results were statistically significant if the $|t|$ or $|z|$ values were ≥2.

Modulation of landing positions by trial number was tested with Generalised Additive Mixed Models (GAMMs) (Baayen et al., 2017; Sóskuthy, 2017; Wieling, 2018; Wood, 2017). GAMMs are an extension of GLMMs where part of the predictors are specified as smooths. These smooths represent the weighted sum of a number of base functions (Baayen et al., 2017). In this model, cubic regression splines were used as the base functions (Sóskuthy, 2017). The addition of smooths makes GAMMs well-suited to model temporally-correlated data, particularly if they exhibit a potentially non-linear relationship. Smooth terms were added for the effect of trial number within blocks, as well as for the by-subject and by-item random intercepts and random slopes. The GAMM models were fit with the "mgcv" v.1.8-26 R package (Wood, 2017) and visualised with the "itsadug" v.2.3 R package (van Rij et al., 2017). The remaining graphs were generated with ggplot2 (Wickham, 2016).

## 3. Results

The mean comprehension accuracy was 97.9% (SD= 14.4%), indicating that participants understood the sentences. There were no significant differences in comprehension accuracy across the conditions (all $|z| \leq 0.94$). The data were pre-processed manually with EyeDoctor (Stracuzzi & Kinsey, 2009) to align the vertical position of fixations whenever necessary[5]. During the pre-processing, 0.05% of trials were removed due to tracking loss. A further 0.03% were excluded as participants made no return-sweeps on that trial. Additionally, 4.44% of trials were discarded due to blinks occurring on return-sweeps or immediately before or after a return-sweep. Fixations shorter than 80 ms that

---

[5] This was a data quality monitoring procedure in which fixations were manually reviewed for cases where the fixation occurred between lines or a few pixels below or above the text area. In such cases, the fixation was realigned to the line of the adjacent fixations. Overall, 2624 out of all 162871 fixations were re-aligned (0.016 %).

occurred within 14 pixels (the mean of the two font size conditions) of a temporally adjacent fixation were merged into that adjacent fixation (e.g., see Perea & Acha, 2009; Rayner,

Warren, Juhasz, & Liversedge, 2004). However, the results did not change when these fixations were not merged. Any remaining fixations less than 80 ms were discarded[6].

Fixations longer than 1000 ms and their adjacent saccades were removed as outliers (0.03%).

This left 95.45% of the data for analysis (6109 trials). Descriptive statistics are shown in

Table 1.

Table 1

*Mean Descriptive Statistics of Return-sweep Saccade Measures in the Experiment (SDs in Parenthesis)*

| Line length | Font size | Landing position (deg) | Under-sweep probability |
|---|---|---|---|
| Short | Small | 1.05 (1.17) | 0.51 (0.50) |
| Short | Large | 1.38 (1.33) | 0.44 (0.50) |
| Long | Small | 1.99 (1.57) | 0.79 (0.41) |
| Long | Large | 2.34 (1.82) | 0.71 (0.45) |

*Note*: Landing position is measured in degrees of visual angle.

### 3.1. Landing Position

The landing position results are illustrated in Figure 2a and the LMM analysis is shown in Table 2. There was main effect of line length, indicating that participants landed further to the right in the long compared to the short line condition ($d$= 0.53). Additionally, landing positions shifted further to the right in the large compared to the small font condition ($d$= 0.21). This indicates that character size information is used when programming return-sweeps. The returns-sweep launch position effect was also significant, indicating that return-sweeps that were launched further away from the end of the first line landed closer to the left margin. Critically, however, there was no two-way interaction between font size and line length. Therefore, the use of character information for saccade targeting did not depend on the visual acuity of the target.

---

[6] There were more fixations less than 80 ms for under-sweep (n= 40) compared to accurate-sweep cases (n= 4), $\chi^2(1)$= 29.455, $p$= 5.7x10$^{-8}$. However, keeping these fixations in the data did not change the main results or the conclusions from the analyses.

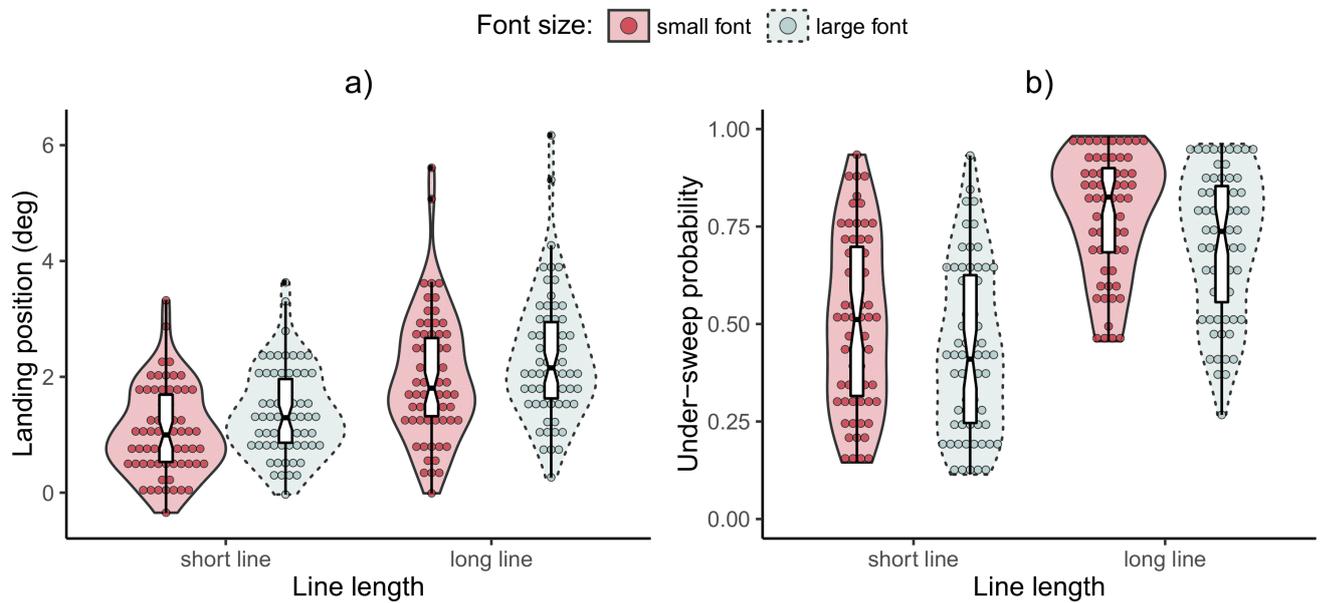

*Figure 2*. Box plots and probability densities for the landing position (**a**) and under-sweep probability (**b**) measures. Landing position was measured in degrees of visual angle relative to the start of the line. Dots represent the mean value for each subject, as estimated by the (G)LMM model. The central mark on each boxplot shows the median.

Table 2

*(G)LMM Results for Landing Position in Degrees of Visual Angle and Under-sweep*

*Probability as a Function of Font Size, Line Length, and Return-sweep Launch Position*

| Fixed effects | Landing position (deg) | | | Under-sweep probability | | |
|---|---|---|---|---|---|---|
| | b | SE | t | b | SE | z |
| Intercept | 1.685 | 0.117 | **14.354** | 0.643 | 0.149 | **4.329** |
| Font size | 0.179 | 0.015 | **12.157** | -0.189 | 0.032 | **-5.934** |
| Line length | 0.464 | 0.033 | **13.943** | 0.751 | 0.057 | **13.211** |
| Launch position | -0.05 | 0.016 | **-3.151** | -0.295 | 0.036 | **-8.111** |
| Font size x Line length | 0.01 | 0.015 | 0.655 | -0.042 | 0.032 | -1.327 |
| Font size x Launch position | 0.032 | 0.015 | **2.144** | 0.013 | 0.034 | 0.396 |
| Line length x Launch position | 0.006 | 0.016 | 0.377 | < 0.01 | 0.035 | -0.007 |
| Font size x Line length x Launch position | 0.046 | 0.015 | **3.02** | 0.011 | 0.034 | 0.316 |
| Random effects | Var. | SD | Corr. | Var. | SD | Corr. |
| Intercept (items) | 0.039 | 0.197 | - | 0.171 | 0.413 | - |
| Intercept (subjects) | 0.842 | 0.917 | - | 1.220 | 1.104 | - |
| Line length slope (items) | 0.006 | 0.080 | 0.93 | 0.026 | 0.162 | 0.22 |
| Line length slope (subjects) | 0.052 | 0.229 | 0.73 | 0.107 | 0.327 | 0.11 |
| Residual | 1.276 | 1.129 | - | - | - | - |

*Note*: statistically significant t-/ z-values are formatted in bold. Return-sweep launch position was centred with a mean of 0.

Furthermore, there was a significant two-way interaction between font size and return-sweep launch position. The main effect of launch position was less pronounced in the large compared to the small font condition. Therefore, return-sweep launch position exerted less of an influence on landing positions when there were fewer, but bigger, characters on the line. Interestingly, the three-way interaction between font size, line length and return-sweep launch position also reached significance. As Figure 3 illustrates, in the small font condition, return-sweep launch positions that were closer to the left margin resulted in return-sweeps that landed closer to the left margin for both line length conditions. However, in the large font condition, the return-sweep launch position effect went in the opposite direction for the two line length conditions. While the same effect was observed for short lines, the trend went in the opposite direction for long lines. That is, launch positions that were closer to the left margin paradoxically led to landing positions further away from that margin. Therefore, the return-sweep launch position effect was less pronounced in the large font sentences because the two line length conditions were largely cancelling each other out.

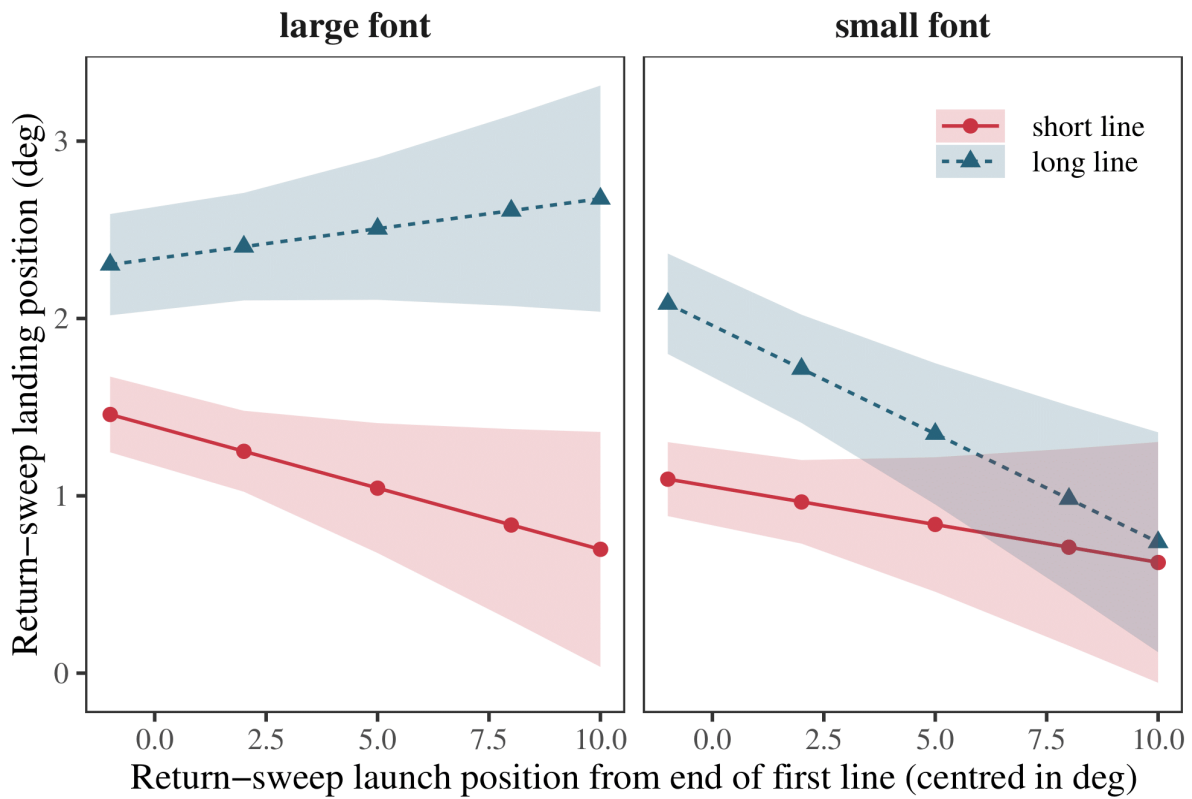

*Figure 3.* Three-way interaction between font size, line length and return-sweep launch position in the landing position model. Landing position was measured in degrees relative to the start of the line and return-sweep launch position was measured in degrees relative to the end of the previous line. Greater return-sweep launch position numbers thus indicate greater distance from the end of the first line and therefore correspond to a shorter distance to the beginning of the next line. Shading indicates ±1 SE. The fitted values from the model were extracted with the "effects" R package v.4.1 (Fox & Hong, 2009).

## 3.2. Modulation of Landing Position by Trial Number

One question of interest was how trial number within the font size blocks may influence participants' landing positions. If landing positions are modulated by trial number, this would suggest that participants gradually learn to adjust their targeting decisions based on increased exposure to a given font size. A GAMM model was fit comparing the two font size conditions. Two models were run, one for each of the two block orders in the experiment: small font first – large font second and large font first – small font second (henceforth, "small - large" and "large - small" order). The results are visualised in Figure 4.

In the small- large block order, the smooth term of trial number was not significant ($edf$= 1.00, $F(1)$= 0.812, $p$= 0.367), indicating that landing positions did not change with trial number in both font size conditions. However, the interaction between trial number and small font was significant ($edf$= 1.000, $F(1)$= 11.494, $p$< 0.001), thus showing that landing positions in the small font condition were modulated by trial number. As Figure 4a shows, this occurred because return-sweep landing positions continuously shifted to the left towards the line margin as participants were exposed to more small font sentences (i.e., as trial number increased). Visually, there was a slight tendency for large font sentences to shift in the opposite direction, but the interaction between trial number and large font was not significant (edf < 0.001, $F(0.00003)$= 0.057, $p$= 0.998). The difference between the small and large font condition (Figure 4c) showed that the font size effect emerged around the 15th trial and continued to increase until the end of the experimental block.

In the large-small block order, a similar pattern of results was obtained. The smooth term for trial number was significant this time (*edf*= 3.002, *F*(3.735)= 2.707, *p*= 0.02668), indicating a change in landing positions with trial number in both font size conditions. As Figure 4b shows, landing positions in the small font condition gradually shifted to the left with increasing trial number, whereas landing positions in the large font condition generally shifted to the right. The interaction between trial number and small font did not reach significance this time, but it was in the same direction as the same interaction in the small-large block order (*edf*= 1.539, *F*(1.883)= 2.951, *p*= 0.083). The interaction between trial number and the large font condition was also not significant (*edf*= <0.0002, *F*(0.00004)= 0.028, *p*= 0.999). The difference between the small and large font conditions (Figure 4d) indicated that the font size effect emerged around the 6[th] trial and again continued to increase throughout the experimental block. In summary, the GAMM results provided at least some evidence that landing positions were modulated by greater exposure to a given font, because participants continued to adjust their landing positions as they read more sentences of the same font. Nevertheless, it should be mentioned that this effect was mostly driven by changes in the small font condition.

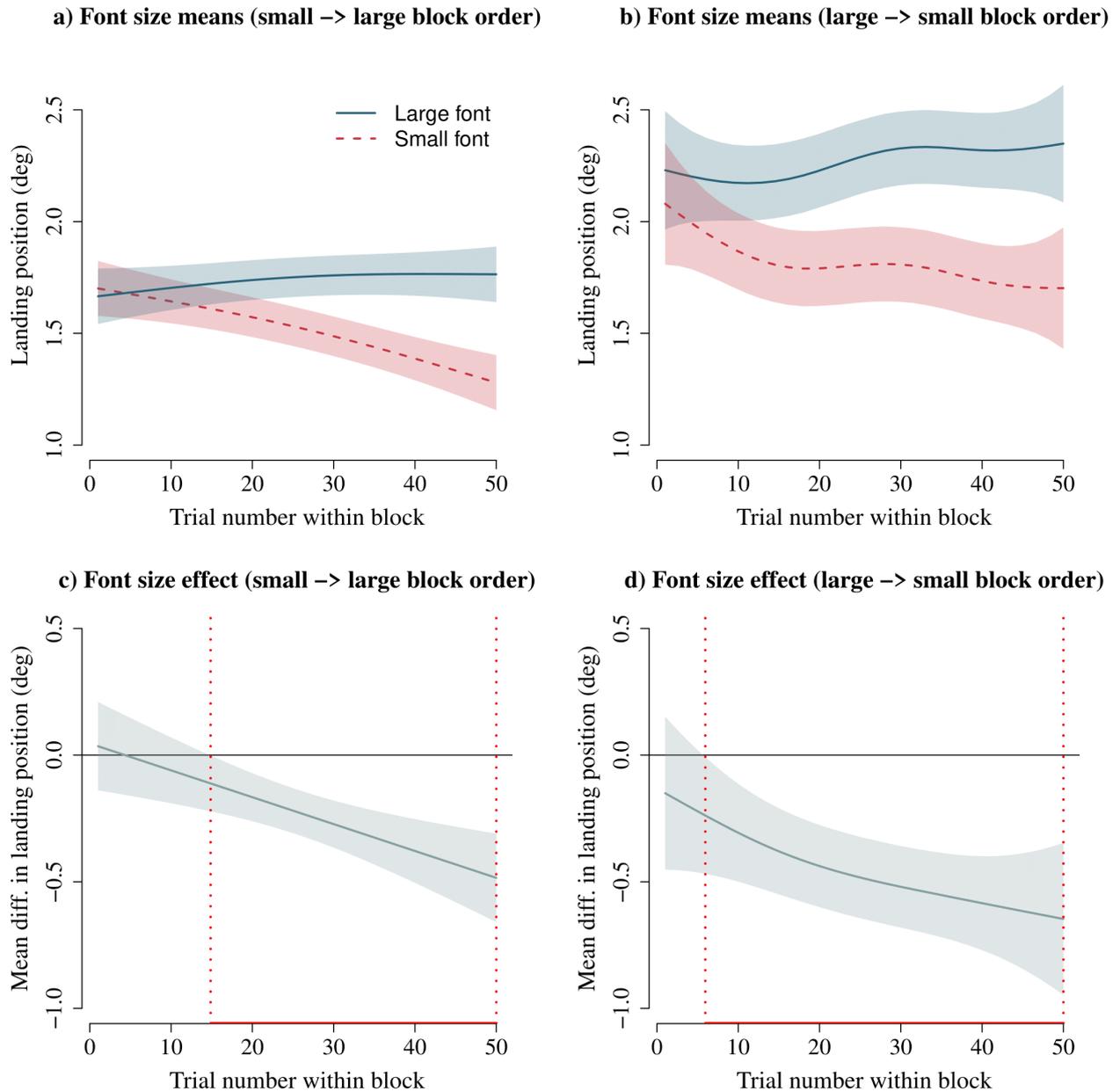

*Figure 4*. Modulation of return-sweep landing positions by trial number in the experiment. Plotted are the estimated slope smooths from the GAMM model for the large and small font conditions in the small –> large font block order (**a**) and in the large –> small font block order (**b**). The two block orders were counterbalanced across participants. The estimated mean difference of the small compared to the large font condition is shown in (**c**) for participants who saw the small block first and large block second and in (**d**) for participants who saw the large block first and the small block second. Shading indicates the 95% confidence interval. The mean difference between the two font conditions in panels (**c**) and (**d**) is significant when 95% confidence interval excludes 0 (denoted by vertical dotted lines).

### 3.3. Under-sweep Probability

The under-sweep probability results are illustrated in Figure 2b and the GLMM model results are presented in Table 2. There was a main effect of font size, indicating that under-sweep probability decreased in the large compared to the small font condition ($d$= -0.15). This shows that readers were less likely to make an under-sweep fixation in the large font condition. Additionally, under-sweep probability increased in the long line compared to the short line condition ($d$= 0.59). Therefore, readers were more likely to make an under-sweep fixation in the long line condition, presumably because return-sweeps landed further to the right of the line margin compared to the short line condition. There was also a main effect of return-sweep launch position, which was due to greater probability of making an under-sweep fixation when the launch position was further from the left margin to begin with. Finally, the interaction between font size and line length was not significant—the font size effect was not modulated by the length of the previous line.

### 4. Discussion

We examined whether return-sweep saccades are guided by character information and whether visual acuity constraints modulate the use of such information. The key findings can be summarised as follows. First, consistent with Hofmeister (1998), the larger font led to return-sweeps landing further to the right of the line start and a smaller probability of making an under-sweep fixation compared to the smaller font. This clearly suggests that character information is used in return-sweep planning. Second, when return-sweeps were launched from a longer line, landing positions also shifted to the right and the probability of making an under-sweep fixation increased. This likely reflects the increase in saccadic error, also replicating previous results (Hofmeister et al., 1999). Third, visual acuity at the saccade target did not modulate the use of character information, as indicated by the lack of a font size by line length interaction. This suggests that the use of font size information does not depend on how far into the periphery the saccade target is, at least within the range of 16-26º. Finally, there was at least some evidence to suggest that readers dynamically modulate their return-sweep landing positions with increased exposure to a given font size, although this was mostly evident in the smaller font block. We hypothesise that with the smaller font block, landing sites which were too far to the right would be less optimal and likely require a leftward saccade in order to process the smaller characters to the left of fixation. However, with the larger font, there may be less pressure for readers to adjust their targeting strategy because there will be fewer and larger letters to the left.

While return-sweeps traverse a much larger distance than intra-line saccades, the present data clearly suggest that these longer saccades are guided by font size information. Therefore, the present results do not support a saccade targeting strategy in which readers always aim for the same physical location on the next line while ignoring letter size information. Rather, readers take into account the formatting of the text and aim for a location on the next line that allows for efficient processing of letters based on their size. Additionally, the use of character information did not depend on visual acuity at the saccade target as the font size effect on landing positions was not modulated by the length of the previous line. This suggests that font size information is used as a global targeting cue because it does not depend on how well readers can perceive words at the beginning of the next line. Interestingly, return-sweeps are also influenced by letter spacing in a similar way to font size- when the spacing between letters increases, landing positions also shift to the right (Hofmeister, 1998, Experiment 3). Therefore, even though visual acuity at the line start is usually limited, readers use global text formatting information to help them land in a more optimal viewing position. It may be that global text information available in foveal vision is extrapolated in order to assist in the programming of return-sweeps, which have targets too distant for accurate bottom up letter encoding.

The preference to land in a position optimized for the size of letters is interesting when one considers the fact that return-sweeps do not appear to target the OVP of the line-initial word (Slattery & Vasilev, 2019). Because return-sweeps are typically launched at least 10º away from the target, accurately aiming for the first word's OVP may not be a feasible strategy. Rather, by aiming for a location relative to the left margin, readers may attempt to land in a position that leaves a few characters to the left of fixation. Consequently, the landing position in visual angle would be a function of font size—with larger fonts, this position would shift to the right to compensate for the bigger letters. This may be advantageous as the few characters to the left of fixation will usually fall within foveal vision and readers may be able to process line-initial information more optimally than if they had landed at the left margin itself. Additionally, landing a few characters from the line start may minimize the probability of overshooting the line start, which could reduce the overall flight time (Harris, 1995) or energy expenditure (Becker, 1989) of such eye-movements. Therefore, it may be more optimal to attempt to land a few characters from the line start.

Indeed, our analysis of trial number effects adds some preliminary support to this explanation. Within a block, readers appear to gradually adjust the target location of their return-sweeps so that they land around the 5th character on the line in both the small and large font size conditions. However, this gradual shift was only statistically significant in the small font block. As already mentioned, this may be due to a larger cost of landing too far from the line start when the characters on the line are smaller. However, an alternative explanation could be that our participants were more accustomed to reading text closer in size to that of our large font condition, so such adjustment may be less needed with larger font sentences.

There was one unexpected effect in the landing position analysis. We found evidence of a three-way interaction between font size, line length, and return-sweep launch position which also drove a two-way interaction between font size and launch position. With the small font, landing positions shifted closer to the start of the new line as launch positions were further from the end of the prior line and this effect was more pronounced for the long line condition. However, for the large font, the effect of launch position was similar to the small font with the short lines but went in the opposite direction for the long lines (see Figure 3). This isn't the first time that return-sweep launch position has been involved with an unpredicted interaction in an analysis of return-sweep landing position data. Recently, Slattery and Vasilev (2019) reported that return-sweep launch position interacted with the length of line-initial words in predicting landing positions. Clearly, more research is needed to understand how launch position influences return-sweep targeting.

One interesting finding was that under-sweep probability was smaller when the font size was larger, presumably because there was less character information to the left of fixation to process, as suggested by Hofmeister (1998). This result indicates that under-sweeps are not programmed solely as a function of the distance in visual angle between the landing position and the left margin. Clearly, the present data indicate that readers use character information at least to some degree to decide whether a correction is needed. However, it is important to note that under-sweep probability is not solely based on character information. This was because there was no interaction between font size and line length in the under-sweep probability data. Because return-sweeps launched from a longer line will undershoot the line start to a greater extent compared to those launched from a short line, the difference in the number of characters to the left of fixation between the two font size conditions was 0.06 characters in the short line condition and 0.81 characters in the long line condition. Therefore, if the probability of making an under-sweep is based only on character information, then we should have found an interaction between these two variables. However, the difference in under-sweep probability was nearly identical in the two conditions (7% short lines, 8% long lines). Moreover, the font size difference in number of characters to the left of the landing site in the short line condition was so small that there should have been no effect of under-sweep probability according to Hofmeister's speculation.

One alternative explanation for the font size effect in under-sweep probability is that the landing of return-sweeps may be more accurate in the large font condition with respect to their intended location, so fewer corrections may be needed. To test for this possibility, we compared the variance of return-sweep landing positions in the two font size conditions to test whether they come from the same population. A Fligner-Killeen test found no significant difference between the small and large font conditions, $\chi^2(1)= 0.772$, $p= 0.379$, thus suggesting that their variances are homogenous and come from the same distribution. Therefore, the data did not support the view that the font size effect in under-sweep probability may emerge from less variability of return-sweep landing positions in the larger font condition.

While the effect of font size was found in both landing positions and under-sweep probability, it is not clear whether its origin is the same. Because return-sweeps are programmed at the end of the previous line, the landing position effect must originate prior to the execution of the return-sweep. However, the corrective regression is executed after readers have already landed on the next line, which means that they will have a higher resolution view of the line start. This higher resolution view may provide readers with more confidence that they can process the letters to the left of fixation when the font is larger. Still, it is not known whether the influence of font size on under-sweep probability is based on visual feedback once readers have landed on the next line or whether it originates prior to the return-sweep saccade.

There has been a discussion in the literature about whether corrective regressions are pre-programmed before the main saccade (Barnes & Gresty, 1973; Becker, 1972, 1976; Becker & Fuchs, 1969; Shebilske, 1976) or whether they are based on visual feedback following the main saccade (Prablanc et al., 1978; Prablanc & Jeannerod, 1975). Because corrective regressions can occur even in the dark without any visual feedback (e.g., Barnes & Gresty, 1973; Becker & Fuchs, 1969), it has been suggested that they may come "pre-packaged" with the main saccade to save time (Becker & Fuchs, 1969). However, there is also some evidence showing that corrective regressions do not occur in the absence of visual feedback (Prablanc & Jeannerod, 1975). While this discrepancy could be partly due to methodological differences (Becker, 1976), it is important to note that the two viewpoints are not mutually exclusive. For example, recent evidence has suggested that, when the main saccade undershoots the target by more than 10%, a corrective regression is likely to occur regardless of whether visual feedback was present or not (Tian et al., 2013). Therefore, visual feedback may play a stronger role when the undershoot error is smaller.

Currently, little is known about how readers program a leftward saccade following an under-sweep fixation. Therefore, it is not clear whether the font size effect in under-sweep probability arises from visual feedback on the next line or whether it is "pre-packaged" with the main saccade based on the general expectation of larger letters in the text. Nevertheless, regardless of when this influence occurs, the present data indicate that readers use global text characteristics such as font size to determine if a correction is needed. In summary, the present study suggests that font size information is used as a global saccade targeting cue to help readers land in a more optimal viewing position at the start of the new line.

**Acknowledgments**

V.I.A. and C.L. contributed equally to this work. We thank Françoise Vitu and 1 anonymous reviewer for their valuable comments on a previous version of this manuscript.

**Funding**

M.R.V. was supported by a Postdoctoral fellowship from Bournemouth University. This work was made possible in part by a research grant from the Microsoft Corporation to T.J.S. which was matched by Bournemouth University to fund C.L. and V.I.A.'s PhD studies.

**Data Statement**

All data files, analysis scripts, preregistered hypotheses, and materials are available at: https://doi.org/10.17605/OSF.IO/PQHAF

**Declarations of Interest:** None

**CRediT statement**

Conceptualization: M.R.V., M.B., T.J.S. Methodology and Software: M.R.V., V.I.A., C.L., T.J.S. Investigation: C.L., V.I.A. Formal Analysis and Data Curation: M.R.V., V.I.A., C.L., T.J.S. Writing - Original Draft: M.R.V., V.I.A., C.L., M.B., T.J.S. Writing - Review & Editing: M.R.V., T.J.S. All authors approved the final article. Supervision: M.R.V, T.J.S. Project administration: M.R.V. Funding acquisition: T.J.S.

**Supplemental Materials**

**Saccade Length: Comparison between Intra-line and Return-sweep Saccades**

We compared intra-line and return-sweep saccade length in visual angle to test how font size affects both types of saccades. Because reading models assume that intra-line saccades are targeted towards the words' OVP (e.g., Engbert et al., 2005; Reichle et al., 1998), the progressive saccade length in visual angle should become larger with increasing font size (Slattery & Rayner, 2013; Slattery et al., 2016). On the other hand, the length of return-sweep saccades should either be shorter when the font is larger if readers use font size information for saccade targeting, or there should be no difference in return-sweep saccade lengths between the font size conditions if readers simply target the left line margin and ignore font size information. This should therefore result in an interaction between font size and saccade type (intra-line vs. return-sweep).

Return-sweeps can be considered as progressive saccades because they bring gaze towards unexplored parts of the text on the next line. Therefore, the main analysis focused on comparing return-sweeps to forward intra-line saccades. However, for completeness, we also report a comparison to regressive intra-line saccades. In the LMM models, saccade type was entered as a factor in addition to font size and line length (intra-line saccades were coded as -1 and return-sweep saccades were coded as 1). Line length was included as a predictor for completeness, but the interaction between font size and saccade type (intra-line vs. return-sweep) was of main theoretical importance[7].

Descriptive statistics are shown in Table S1 and the LMM results are presented in Table S2 for forward intra-line saccades and Table S3 for regressive intra-line saccades. In the main analysis comparing forward intra-line saccades to return-sweeps, there was a main effect of saccade type, indicating that return-sweeps travelled farther in visual angle compared to intra-line saccades. Additionally, saccades were significantly longer when the first line of text was also longer. This effect was largely due to the increase in saccade length for return-sweeps with longer lines, which is shown by the robust line length by saccade type interaction. Crucially, consistent with our hypothesis above, there was a significant interaction between font size and saccade type. This occurred because forward intra-line saccade lengths increased as the font size increased but return-sweep saccade lengths decreased as font size increased. In both cases, this change reflects a change in saccade amplitude to accommodate the different size of letters.

---

[7] The pre-registration also included return-sweep launch position as a covariate. However, as this is not of main theoretical interest, the covariate was not included.

Table S1

*Mean Descriptive Statistics for Saccade Length in the Experiment (SDs in Parenthesis)*

| Line length | Font size | Forward intra-line saccade length | Regressive intra-line saccade length | Return-sweep saccade length |
|---|---|---|---|---|
| Short | Small | 2.53 (1.43) | 2.80 (2.65) | 12.7 (2.16) |
| Short | Large | 3.05 (1.60) | 3.41 (3.02) | 11.9 (2.33) |
| Long | Small | 2.70 (1.54) | 2.91 (3.30) | 22.4 (2.78) |
| Long | Large | 3.26 (1.78) | 3.57 (3.46) | 21.6 (2.93) |

*Note*: Saccade length was measured in degrees of visual angle.

The secondary comparison of regressive intra-line saccades and return-sweeps revealed the same pattern of results (see Table S3). The only difference from the analysis of forward intra-line saccades was that the main effect of font size did not reach statistical significance (but was still the same direction). However, the critical interaction between font size and saccade type was again highly significant. This interaction reflected the same relationship as the model with forward intra-line saccades: while regressive saccade length increased in the larger font condition, return-sweep saccade length decreased. In summary, the font size manipulation affected both intra-line and return-sweep saccade amplitudes.

Table S2

*LMM Results for the Comparison between Forward Intra-line Saccade Length and Return-sweep Saccade Length*

| Fixed Effects | b | SE | t |
|---|---|---|---|
| Intercept | 7.566 | 0.05 | **151.63** |
| Font size | -0.075 | 0.019 | **-3.986** |
| Line length | 4.944 | 0.022 | **225.13** |
| Saccade type | 4.721 | 0.016 | **303.11** |
| Font size x Line length | 0.007 | 0.022 | 0.30 |
| Font size x Saccade type | -0.344 | 0.016 | **-22.13** |
| Line length x Saccade type | 4.759 | 0.022 | **216.77** |
| Font size x Line length x Saccade type | -0.009 | 0.022 | -0.40 |
| Random Effects | Var. | SD | Corr. |
| Intercept (items) | 0.0158 | 0.1255 | - |
| Intercept (subjects) | 0.1337 | 0.3657 | - |
| Font size slope (subjects) | 0.0072 | 0.0847 | 0.42 |
| Residual | 2.71369 | 1.6473 | - |

*Note*: statistically significant *t*-values are formatted in bold.

Table S3

*LMM Results for the Comparison between Regressive Intra-line Saccade Length and Return-sweep Saccade Length*

| Fixed Effects | b | SE | t |
|---|---|---|---|
| Intercept | 7.688 | 0.068 | **112.28** |
| Font size | -0.061 | 0.033 | -1.821 |
| Line length | 4.922 | 0.043 | **113.96** |
| Saccade type | 4.601 | 0.031 | **147.26** |
| Font size x Line length | 0.005 | 0.043 | 0.121 |
| Font size x Saccade type | -0.357 | 0.031 | **-11.53** |
| Line length x Saccade type | 4.782 | 0.043 | **110.75** |
| Font size x Line length x Saccade type | -0.007 | 0.043 | -0.167 |
| Random Effects | Var. | SD | Corr. |
| Intercept (items) | 0.0661 | 0.2570 | - |
| Intercept (subjects) | 0.1950 | 0.4416 | - |
| Font size slope (subjects) | 0.0102 | 0.1012 | 0.55 |
| Residual | 8.9875 | 2.9979 | - |

*Note*: statistically significant *t*-values are formatted in bold.

**Modulation of Forward Saccade Length by Trial Number**

Similar to the analysis of return-sweep landing positions in the main text, a GAMM model was fit to analyse how forward saccade length changed as a function of trial number. The results are illustrated in Figure S1. In the small – large font block order, neither the smooth term of trial order ($edf$= 1.004, $F(1.005)$= 0.126, $p$= 0.7256), nor its interactions with the small ($edf$= 1.000, $F(1)$= 0.103, $p$= 0.749) or large font condition ($edf$= 1.230, $F(1.449)$= 2.023, $p$= 0.303) were statistically significant. This suggests that the effect of font size on intra-line saccade lengths was generally not modulated by trial number. In the large – small font block order, the smooth term of trial number reached significance ($edf$= 2.269, $F(2.469)$=

6.336, $p$= 0.0012), but the interactions between trial number and the small font (*edf*= 0.0005, $F$(0.0007)= 0.152, $p$= 0.992), and trial number and the large font (edf= 1.001, F(1.001)= 0.321, $p$= 0.571) were not significant. As Figure S1b shows, there was tendency for saccade lengths in both font size conditions to slightly increase with trial number. Nevertheless, the mean difference between the two font size conditions remained relatively linear and did not how much change as a function of trial number (see Figure S1c-d). To summarise, the effect of font size on forward intra-line saccades was significant from the initial trials of each block and remained so with only a limited influence of trial number throughout the blocks. This is consistent with an immediate adjustment to saccade targeting based on parafoveally available letter size information.

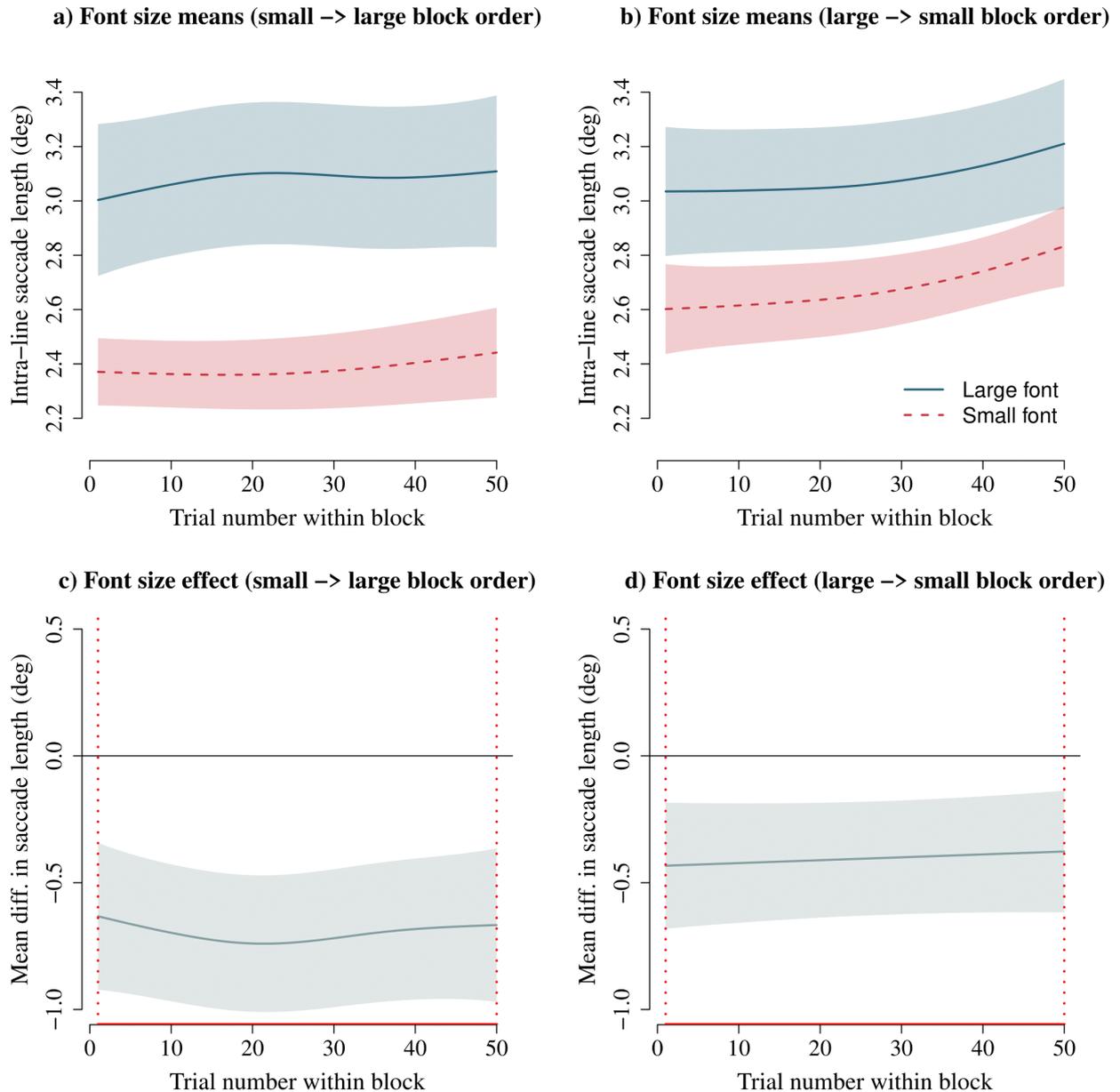

*Figure S1*. Modulation of forward intra-line saccade length by trial number in the experiment. Plotted are the estimated slope smooths from the GAMM model for the large and small font conditions in the small –> large font block order (**a**) and in the large –> small font block order (**b**). The two block orders were counterbalanced across participants. The estimated mean difference of the small compared to the large font condition is shown in (**c**) for participants who saw the small block first and large block second and in (**d**) for participants who saw the large block first and the small block second. Shading indicates the 95% confidence interval. The mean difference between the two font conditions in panels (**c**) and (**d**) is significant when 95% confidence interval excludes 0 (denoted by vertical dotted lines).

**Modulation of Return-sweep Saccade Length by Trial Number**

We conducted the same analysis of return-sweep saccade lengths by fitting a GAMM model to analyse how return-sweep saccade length changed as a function of trial number. The results are illustrated in Figure S2. The pattern of results was very similar to the analysis of landing positions reported in the main text. In the small- large block order, there was a tendency for return-sweep saccade length to increase with trial number in the small font condition and to decrease in the large font condition. Nevertheless, neither the smooth term of trial order (*edf*= 1.000, *F*(1)= 1.189, *p*= 0.276) nor its interaction with the small font (edf= 0.00001, F(0.00002)= 0.101, *p*= 0.998) or large font conditions (edf= 1.130, F(1.248)= 3.151, *p*= 0.086) reached significance. The difference between the small and large font conditions became significant around the 14th trial and continued to increase as the experiment progressed.

In the large-small block order, there was a similar trend for the small font condition, whereas the large font condition remained relatively constant with increasing trial number. Again, neither the smooth term of trial order (*edf*= 1, *F*(1)= 0.004, *p*= 0.948), nor it's interaction with the small font (*edf*= 1.910, *F*(2.383)= 0.978, *p*= 0.499) or the large font condition (*edf*= 0.00002, *F*(0.00004)= 0.042, *p*= 0.999) was significant. The mean difference between the small and large font condition because significant around the 10th trial and increased until around the 20th trial before plateauing. In summary, the pattern of results from the returns-sweep length analysis was generally similar to that of the landing position analysis, despite the smooth terms not being significant. In general, there was greater uncertainty around the estimates for return-sweep saccade lengths, which could have contributed to the lack of significance. Additionally, unlike landing positions, returns-sweep saccade lengths are also affected by the launch position of the return-sweep saccade. Nevertheless, the mean difference between the two font size conditions resembled the landing position analysis in the sense that the effect was not immediately significant (as in the intra-line saccade analysis above), but only reached significance around the 10-14[th] trial. Additionally, unlike the intra-line saccade lengths (which showed a largely consistent font size difference across trials), the font size difference for return-sweeps continued to increase across trials to a greater extent.

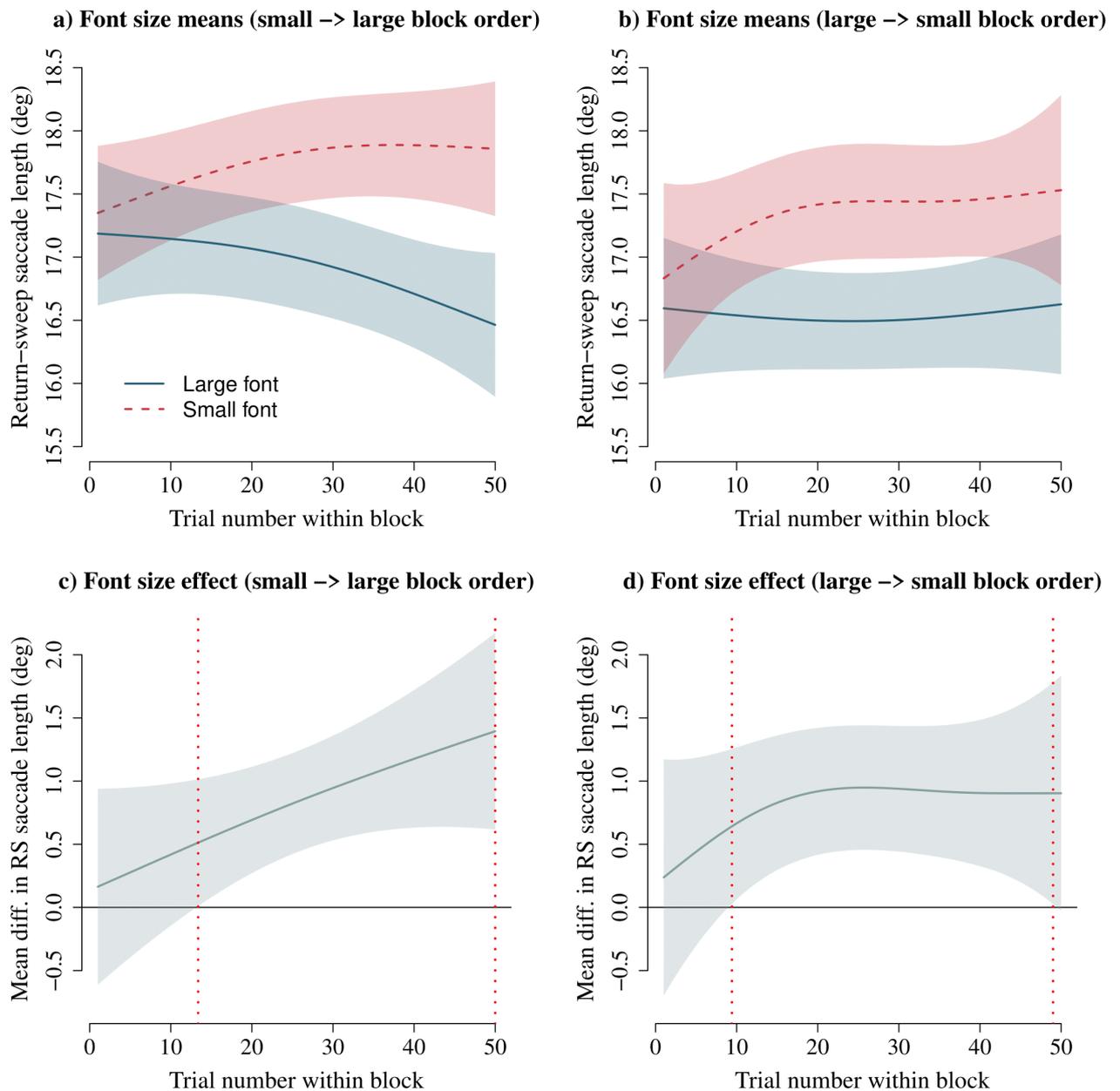

*Figure S2*. Modulation of return-sweep saccade length by trial number in the experiment. Plotted are the estimated slope smooths from the GAMM model for the large and small font

conditions in the small –> large font block order (**a**) and in the large –> small font block order (**b**). The two block orders were counterbalanced across participants. The estimated mean difference of the small compared to the large font condition is shown in (**c**) for participants who saw the small block first and large block second and in (**d**) for participants who saw the large block first and the small block second. Shading indicates the 95% confidence interval. The mean difference between the two font conditions in panels (**c**) and (**d**) is significant when 95% confidence interval excludes 0 (denoted by vertical dotted lines).